\documentclass[letterpaper]{article} % DO NOT CHANGE THIS
\usepackage{aaai24}  % DO NOT CHANGE THIS
\usepackage{times}  % DO NOT CHANGE THIS
\usepackage{helvet}  % DO NOT CHANGE THIS
\usepackage{courier}  % DO NOT CHANGE THIS
\usepackage[hyphens]{url}  % DO NOT CHANGE THIS
\usepackage{graphicx} % DO NOT CHANGE THIS
\usepackage{natbib}  % DO NOT CHANGE THIS AND DO NOT ADD ANY OPTIONS TO IT
\usepackage{caption} % DO NOT CHANGE THIS AND DO NOT ADD ANY OPTIONS TO IT
\usepackage{algorithm}
\usepackage{multirow}
\usepackage{multicol}
\usepackage{subfigure}
\usepackage{bm}
\usepackage{color}
\usepackage{adjustbox}

\usepackage{enumitem}
\usepackage{graphicx}
\usepackage{mathtools}
\usepackage{graphics}
\usepackage{bbding}
\usepackage{algorithm}
\usepackage{algpseudocode}
\usepackage{booktabs} 
\usepackage{newfloat}
\usepackage{listings}
\DeclareCaptionStyle{ruled}{labelfont=normalfont,labelsep=colon,strut=off} % DO NOT CHANGE THIS
\floatstyle{ruled}
\newfloat{listing}{tb}{lst}{}
\floatname{listing}{Listing}
\title{LLMRec: Benchmarking Large Language Models on Recommendation Task}

\author {
    Junling Liu\textsuperscript{\rm 1},
    Chao Liu\textsuperscript{\rm 2},
    Peilin Zhou\textsuperscript{\rm 3},
    Qichen Ye\textsuperscript{\rm 4},
    Dading Chong\textsuperscript{\rm 4},
    Kang Zhou\textsuperscript{\rm 1},
    Yueqi Xie\textsuperscript{\rm 5},
    Yuwei Cao\textsuperscript{\rm 8},
    Shoujin Wang\textsuperscript{\rm 6},
    Chenyu You\textsuperscript{\rm 7},
    Philip S. Yu\textsuperscript{\rm 8}
}
\affiliations {
    \textsuperscript{\rm 1}Alibaba Group, \textsuperscript{\rm 2}Ant Group, \textsuperscript{\rm 3}The Hong Kong University of Science and Technology (Guangzhou)\\
    \textsuperscript{\rm 4}School of Electronic and Computer Engineering, Peking University\\
    \textsuperscript{\rm 5}The Hong Kong University of Science and Technology\\
    \textsuperscript{\rm 6}Data Science Institute,University of Technology Sydney\\
    \textsuperscript{\rm 7}Department of Electrical Engineering, Yale University\\
    \textsuperscript{\rm 8}Department of Computer Science, University of Illinois Chicago\\
    william.liuj@gmail.com, chaoliu\_pku@pku.edu.cn, zhoupalin@gmail.com, yeeeqichen@pku.edu.cn \\
    1601213984@pku.edu.cn, kangbeyond89@163.com, yxieay@connect.ust.hk, ycao43@uic.edu \\shoujin.wang@mq.edu.au, chenyu.you@yale.edu, psyu@uic.edu
}

\usepackage{bibentry}
\begin{document}
\graphicspath{{./images/}}

\maketitle

\begin{abstract}
Recently, the fast development of Large Language Models (LLMs) such as ChatGPT has significantly advanced NLP tasks by enhancing the capabilities of conversational models. 
However, the application of LLMs in the recommendation domain has not been thoroughly investigated. 
To bridge this gap, we propose \textbf{LLMRec}, a LLM-based recommender system designed for benchmarking LLMs on various recommendation tasks. 
Specifically, we benchmark several popular off-the-shelf LLMs, such as ChatGPT, LLaMA, ChatGLM,
on five recommendation tasks, including rating prediction, sequential recommendation, direct recommendation, explanation generation, and review summarization. Furthermore, we investigate the effectiveness of supervised finetuning to improve LLMs' instruction compliance ability.
The results indicate that LLMs display limited proficiency in several accuracy-based tasks. However, they demonstrate comparable performance to state-of-the-art methods in explainability-based tasks.
We also conduct qualitative evaluations to further evaluate the quality of contents generated by different models, and the results show that LLMs can truly understand the provided information and generate clearer and more reasonable results. 
We aspire that this benchmark will serve as an inspiration for researchers to delve deeper into the potential of LLMs in enhancing recommendation performance. Our codes, processed data and benchmark results are available at \url{https://github.com/williamliujl/LLMRec}.
\end{abstract}
\section{Introduction}
\label{intro}
Deep learning based recommender systems (RSs) have gained significant attention due to their ability to capture complex patterns and relationships embedded in user behavior data. These RSs leverage deep neural networks to first learn informative representations of users and items from user-item interaction data and then make accurate recommendations using these representations. They have shown impressive performance in various web applications, including e-commerce \citep{sun2022revisiting, liu2022ecommerce,tsagkias2021challenges, xie2022decoupled}, video platforms \citep{wei2019mmgcn,zhao2019recommending,papadamou2022just}, news websites \citep{wu2022feedrec,wu2020mind,wu2019npa}, and music streaming services~\citep{kowald2020unfairness,singh2022novel}.
However, most existing deep learning based recommendation methods are task and domain specific.
As a result, they require a large amount of task-specific and in-domain data for training numerous recommendation models that cater to various tasks and application scenarios. Such specialization results in a lack of efficient and effective generalization ability, inhibiting these models from effectively utilizing a vast amount of general knowledge.

At the same time, Large Language Models (LLMs), such as ChatGPT,  have showcased remarkable adaptability, significantly enhancing the performance of downstream Natural Language Processing (NLP) tasks, as a general model.
Even under zero-shot settings, where no additional training data is provided for downstream tasks, LLMs still achieve significant performance across numerous NLP tasks, progressively narrowing the gap with human. 
For instance, \citep{dai2023chataug} employs ChatGPT to enhance text data augmentation by rephrasing sentences. Their findings reveal the potential of ChatGPT in effectively diversifying and expanding the training data. \citep{jiao2023chatgpt} finds ChatGPT surpasses the previously state-of-the-art zero-shot model by a considerable margin in the sentiment analysis task. \citep{bang2023multitask} finds ChatGPT outperforms the previous state-of-the-art zero-shot model by a large margin in the sentiment analysis task. 
To further explore the capabilities of LLMs, existing studies have assessed LLMs (mainly focus on ChatGPT) from various aspects of NLP tasks, including reasoning~\citep{liu2023evaluating}, fairness~\citep{li2023fairness}, and ethical issues~\citep{zhuo2023exploring}. 
However, these studies are still confined to the domain of natural language processing. The performance of LLMs under other data modalities, such as user-item interaction data, has not been thoroughly investigated~\citep{liu2023chatgpt}. Besides, whether LLMs can perform well on classical recommendation tasks remains an open question. 

Motivated by the aforementioned facts, in this paper, we explore the potential of creating an LLM-based recommender system to solve recommendation problems in a unified manner. 
This is especially pertinent given that various tasks within the realm of recommendation inherently involve natural language components, such as explainable recommendations, interest summarization, item descriptions, and more.
Existing recommendation models that utilize language models (LMs) typically confine themselves to smaller-sized LMs and involve a finetuning process. For instance, \citep{zhang2021language} propose to leverage BERT~\citep{bert} and GPT2~\citep{radford2018improving} as recommender systems and utilize prompts to transform the session-based recommendation task into a multi-token cloze task. Furthermore, P5~\citep{geng2022recommendation} utilizes a mixture of prompts from various tasks to assist the T5~\citep{raffel2020exploring} in supporting a broad spectrum of recommendation tasks.
However, these methods are not directly transferable to LLMs due to the high cost of finetuning. Additionally, they do not investigate the effectiveness of LLMs for recommendation.

To fill this research gap, we design an LLM-based recommender system, namely LLMRec, to benchmark the performance of various LLMs (e.g., ChatGPT and ChatGLM~\citep{du2022glm}) on Amazon Beauty dataset.
In LLMRec, each LLM is employed as a general-purpose recommendation model to handle five classical recommendation tasks, including rating prediction, sequential recommendation, direct recommendation, explanation generation, and review summarization. 
By doing so, we want to investigate whether LLM's extensive linguistic and world knowledge acquired from large-scale corpora can be effectively transferred to recommendation scenarios and boost recommmendation performance.
In order to equip LLMs with the capability to tackle recommendation tasks, one of the most crucial steps involves constructing high-quality prompts. 
We devise prompts that can fully utilize the inherent knowledge of the large language model, while simultaneously accommodating various recommendation tasks. Prompt construction plays a pivotal role not only in directly inferring from the large language model but also in fine-tuning PLMs for recommendation tasks.

The main findings of this benchmark are as follows:
\begin{itemize}
    \item The off-the-shelf results indicate limited proficiency in accuracy-based recommendation tasks, possibly due to the lack of extensive exposure to candidate items without task-specific fine-tuning.
    \item Supervised finetuning (SFT) can significantly improves LLMs' instruction compliance ability in recommendation tasks. For instance, ChatGLM-6B with P-tuning is superior to ChatGPT in rating prediction, explanation generation, and review summarization.
    \item For explainable recommendation tasks such as explanation generation and review summarization, objective metrics like BLEU and ROUGE fail to accurately measure the true capability of LLM-based recommender systems. While LLMs receive lower scores on these objective metrics, their generated contents are superior to state-of-the-art methods via qualitative evaluation.
\end{itemize}
\par  

\begin{figure*}[t]
\centering
\includegraphics[width=16cm]{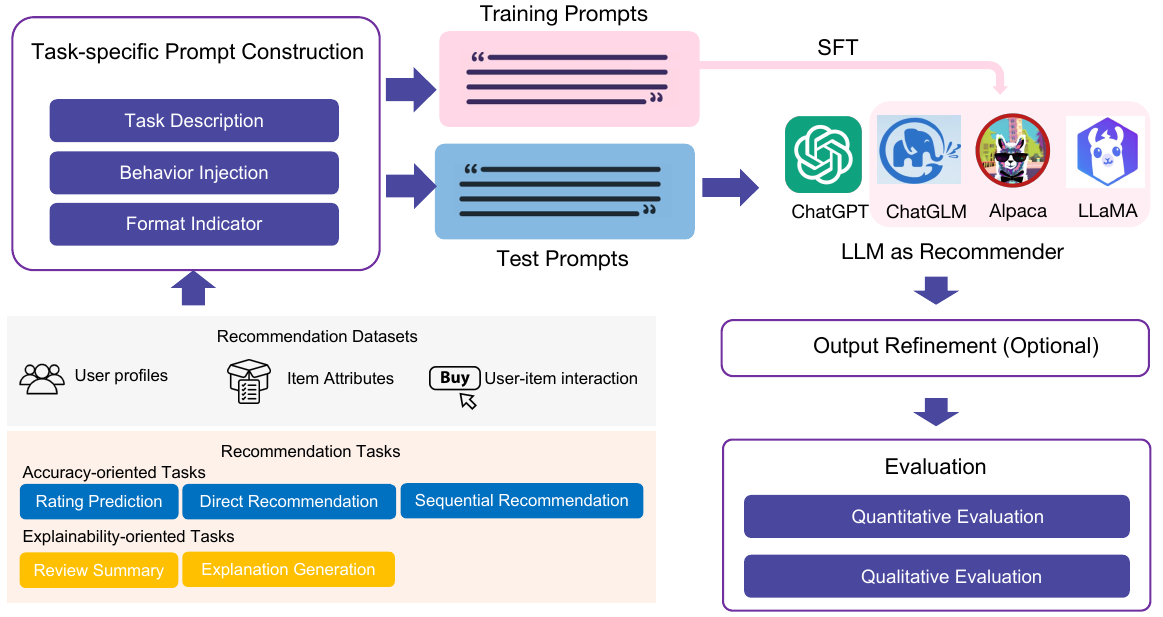}
\caption{The overall architecture of LLMRec.}
\label{fig:workflow}
\end{figure*}
\section{Related Work}
\label{intro}
\par  

\subsection{Language Models}

Language Models (LMs) are a foundational component of natural language processing (NLP) and have been the subject of extensive research over several decades. 
Large Language Models (LLMs) represent a specific category of LMs that harness vast amounts of data and computational resources, enabling them to achieve cutting-edge performance across a wide spectrum of NLP tasks. 
For instance, LLMs excel in machine translation\citep{chen2018best,aharoni2019massively,zeng2022greenplm}, summarization\citep{2017Get,2019Fine}, and dialogue generation\citep{2016A,2016Towards}.
 In the early stage, traditional LMs leverage simple neural architectures including recurrent neural networks (RNNs) and long short-term memory (LSTM) networks to capture dependencies in language\citep{2003Journal,1997Long}. However, traditional neural language models still struggled with capturing the rich semantic and contextual relationships present in natural language. 
 The introduction of the Transformer architecture by \citep{2017Attention} was a major breakthrough in this area. 
 This architecture has been used as the backbone of many successful LLMs, including BERT\citep{bert}, GPT-2\citep{radford2019language}, ChatGPT\citep{gpt4}, ChatGLM \citep{2021GLM}, LLaMA\citep{zhang2023llama}, Alpaca\citep{alpaca}, etc.

\subsection{Large Language Model for Recommendation}
The impressive accomplishments of LLMs have inspired increased attention towards their application in recommender systems, leading to notable advancements in this field.
For instance, LMRecSys~\citep{zhang2021language} tackles zero-shot and data efficiency issues by transforming recommendation tasks into multi-token cloze tasks. P5~\citep{geng2022recommendation} is the first attempt to integrate different recommendation tasks within a shared conditional language generation framework (i.e., T5~\citep{t5}), while M6-Rec~\citep{m6} focuses on building a foundation model to support a wide range of recommendation tasks, including retrieval, ranking, explanation generation, etc. Additionally, Chat-REC~\citep{chat-rec} utilizes ChatGPT to enhance existing recommender models through conversational recommendations, adding interactivity and explainability.
More recently, ~\citep{liu2023chatgpt} conducts a preliminary study that evaluates the potential of ChatGPT across various recommendation tasks.

Our approach differs from the above methods in that we treat LLMs as a self-contained recommender, and based on this, we have designed an LLM-based recommender system called LLMRec. We perform a comprehensive evaluation and comparison of the performance of LLMs on various recommendation tasks. Our objective is to provide researchers with valuable insights to further explore the capabilities of LLMs in recommendation systems.

\section{LLMRec}

\subsection{Task Descriptions}
To conduct a comprehensive test on the recommendation ability of LLMs, we carried out experiments on five tasks: rating prediction, direct recommendation, sequential recommendation, explanation generation, and review summarization. Rating prediction aims to predict the ratings that a user would give to a particular item. 
Sequential recommendation aims to predict a user's next item or action based on their past sequential behavior. 
Direct Recommendation is a type of recommendation system that relies on explicit feedback from users in the form of ratings or reviews. Explanation generation refers to providing users or system designers with explanations to clarify why such items are recommended. Review summarization is primarily used to extract and summarize the content of a series of user reviews, enabling other users to quickly understand the strengths and weaknesses of a product or service. The five tasks mentioned above can be categorized into two groups based on their characteristics: accuracy-based and explainability-based.

\subsection{Overall Architecture}
\label{intro}
\par  
In this paper, we have designed an LLM-based recommender system called LLMRec in order to benchmark the performance of various LLM models on the aforementioned five tasks. The workflow of the proposed recommendation system is illustrated in Fig.\ref{fig:workflow}, which consists of three steps. Firstly, we generate task-specific prompts using task description, behavior injection, and format indicator modules. The task description module is utilized to adapt recommendation tasks to natural language processing tasks. The behavior injection module is designed to assess the impact of Chain-of-Thought (CoT) prompting, which incorporates user-item interaction to aid LLMs in capturing user preferences and needs more effectively. The format indicator serves to constrain the output format, making the recommendation results more comprehensible and assessable. Secondly, we leverage LLM's strong understanding and generation capabilities to directly input the prompt into LLM and obtain corresponding recommendation results. Our LLMRec also supports fine-tuning open-source LLMs using generated prompts to enhance their ability in the recommendation domain. Finally, although LLMs incorporate randomness into the response generation process to ensure diverse results, this randomness can pose challenges in evaluating recommended items. To address this issue, we have developed an output refinement module that checks the format of the recommender's output, corrects any non-compliant results, or feeds them back into LLMs for re-recommendation.

\begin{figure*}[h]
  \centering
  \includegraphics[width=1\textwidth]{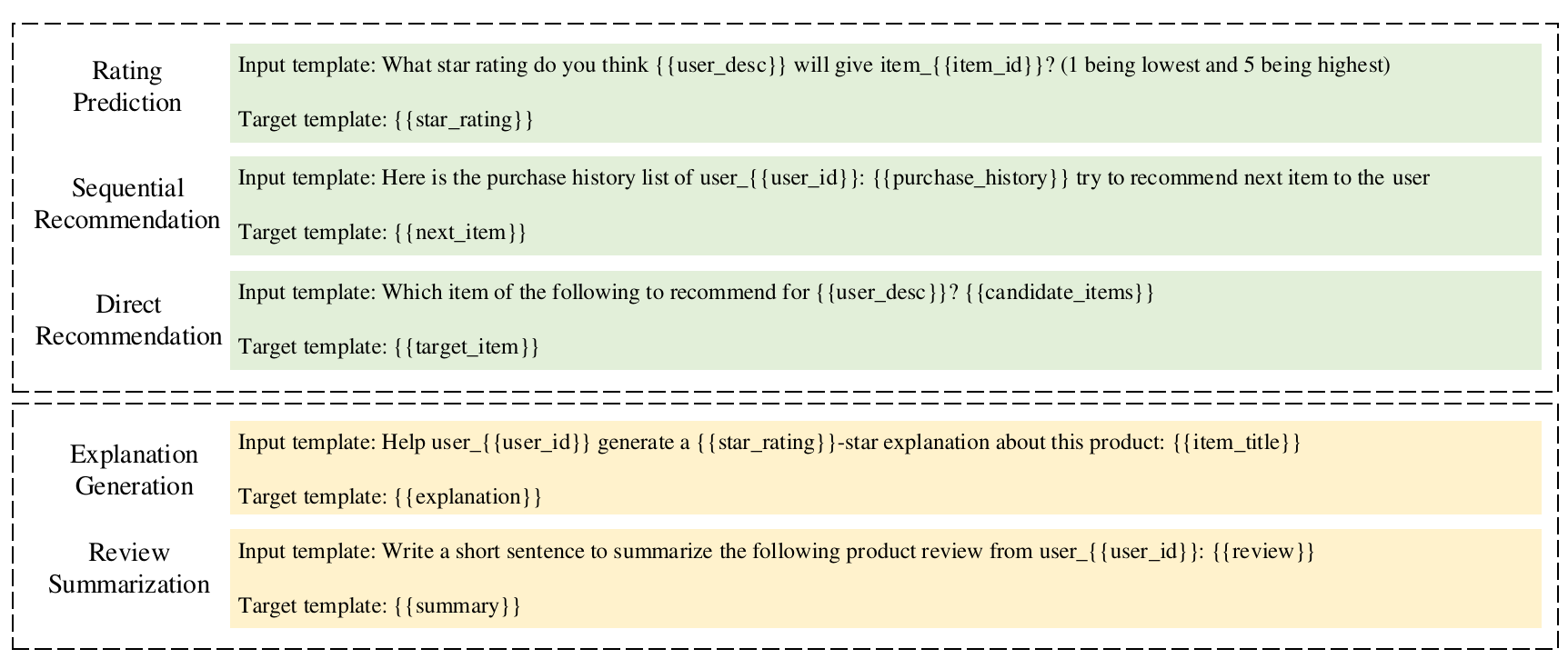}
  \caption{Finetuning prompts for five recommendation tasks.}
  \label{fig:ft-prompts}
\end{figure*}

\subsection{Off-the-shelf LLM as Recommender}
To evaluate the recommendation capabilities of off-the-shelf LLMs in five recommendation tasks, we conducted a benchmarking study on ChatGPT, ChatGLM, LLaMA, and Alpaca using LLMRec. We followed the prompt construction method used in \cite{liu2023chatgpt} and readers can check the details from their paper.

\subsection{Fine-tuned LLM as Recommender}
In order to fully assess the recommendation capabilities of LLMs, we conducted task-specific finetuning on various open-source LLMs, including ChatGLM, LLaMA, and Alpaca, using LLMRec as a foundation. We utilized LLMRec's prompt construction module to produce training data for finetuning. Specifically, as shown in Table~\ref{fig:ft-prompts}, we designed several templates for 5 recommendation tasks according to their characteristics. Due to limited computational resources, we conducted task-specific finetuning using corresponding prompts for each task, instead of jointly training prompts for all tasks as in P5, which aligns more with multi-task finetuning. In the inference process, sequential and direct recommendation tasks typically necessitate an item list as the target output. In that case, for sequential recommendation, beam search algorithms are employed to generate a list of potential next items, which is then evaluated under the all-item setting. In direct recommendation, recommended items are predicted from a candidate set. In this context, beam search is also utilized to decode a list of potential target items with the highest scores, after which evaluations are conducted.

\section{Benchmark}
\subsection{Baselines for recommendation tasks}
Following P5 \citep{geng2022recommendation}, we gather a range of approaches that are representative of various tasks. For rating prediction, we employ MF \citep{koren2009matrix} and MLP \citep{cheng2016wide} as our baselines. For direct recommendation, we use BPR-MF \citep{rendle2012bpr}, BPR-MLP \citep{cheng2016wide} and SimpleX \citep{mao2021simplex} as baselines. For sequential recommendation, we adopt Caser \citep{tang2018personalized}, HGN \citep{ma2019hierarchical}, GRU4Rec \citep{hidasi2015session}, BERT4Rec \citep{sun2019bert4rec}, FDSA \citep{zhang2019feature}, SASRec \citep{kang2018self} and ${\rm S}^3$-Rec \citep{zhou2020s3} as baselines for comparison. For explanation generation, we utilize Attn2Seq \citep{dong2017learning}, NRT \citep{li2017neural} and PETER \citep{li2021personalized} as baselines. For review summarization, we adopt pretrained T0 \citep{sanh2021multitask} and GPT-2 \citep{radford2019language} as baselines.

\subsection{Experimental Setup}
\label{setup}
\subsubsection{Datasets and Metrics.}

Our study involves conducting both quantitative and qualitative evaluations on the real-world Amazon recommendation dataset. This dataset comprises customer review text and associated metadata for products across 29 categories. This paper focuses on evaluating the \textit{Beauty} category.

For rating prediction, we employ Root Mean Square Error (RMSE) and Mean Absolute Error (MAE). For sequential recommendation and direct recommendation, we adopt top-\textit{k} Hit Ratio (HR@\textit{k}), top-\textit{k} Normalized Discounted Cumulative Gain (NDCG@\textit{k}) which are widely used in related works \citep{geng2022recommendation,zhou2020s3}. For explanation generation and review summarization, \textit{n}-gram Bilingual Evaluation Understudy (BLEU-\textit{n}) and 
\textit{n}-gram Recall-Roiented Understudy for Gising Evaluation (ROUGE-\textit{n}) are used for evaluation.

\subsubsection{Implementation Details.}
For off-the-shelf settings, we utilize the gpt-3.5-turbo version of ChatGPT. The rest of the experimental settings remain consistent with \cite{liu2023chatgpt}. For the settings of supervised finetuning, we apply P-tuning V2 \cite{liu2022ptuning} for ChatGLM-6B and LoRa \cite{hu2021lora} for LLaMa-7B as well as Alpaca for efficiently model training. Specifically, we set the prefix sequence length to 128 for P-tuning V2. For LoRa, we set the rank and alpha to 8 and 16, respectively. We train all compared LLMs with batch size of 8 for 10 Epochs across all five tasks. The learning rate was set to 2e-2 for ChatGLM-6B and 3e-4 for LLaMa and Alpaca. In order to trade off the model performance and time consumed, we select one type of prompt template from P5 \cite{geng2022recommendation}'s training data for each task, and evaluate them using the same prompt template. When evaluating models on sequential recommendation and direct recommendation, we perform beam search with beam size 20 to calculate the HR@\textit{k} and NDCG@\textit{k} scores.

\begin{table}[t]
    \centering
    \caption{Performance comparison on explanation generation (\%).}
    \resizebox{\columnwidth}{!}{%
    \begin{tabular}{ccccc}
    \toprule
    Methods
     & BLUE4  & ROUGE1 & ROUGE2  & ROUGEL   \\
    \cmidrule{1-5}
    Attn2Seq    & 0.7889 & 12.6590 & 1.6820 & 9.7481   \\
    NRT    & 0.8295  & 12.7815  & 1.8543  & 9.9477   \\
    PETER   & 1.1541  & 14.8497  & 2.1413  & 11.4143  \\
    P5-B   & 0.9742  & 16.4530  & 1.8858  & 11.8765  \\
    PETER+   & \bf \textbf{3.2606}  & \textbf{25.5541}  &  \textbf{5.9668}  &  \textbf{19.7168}  \\
    ChatGPT  &0.1266 & 6.4060 & 0.3347 & 5.1458\\
    ChatGLM w/o SFT & 0.1318 & 7.8846 & 0.4242 & 4.2388    \\
    LLaMA w/o SFT & 0.0564& 1.0847 &0.1075 &0.6900 \\
    Alpaca w/o SFT &0.2002&5.5061&0.4328&3.3958\\
    \bottomrule
    \end{tabular}}
    \label{tab:explanation}
\end{table}
\begin{table}[t]
    \centering
    \caption{Performance comparison on review summarization (\%).}
    \resizebox{\columnwidth}{!}{%
    \begin{tabular}{ccccc}
    \toprule
    Methods
     & BLUE4  & ROUGE1 & ROUGE2  & ROUGEL   \\
    \cmidrule{1-5}
    T0   &  1.2871  & 1.2750  &  0.3904  &  0.9592  \\
    GPT-2    &  0.5879 &  3.3844 & 0.6756  & 1.3956 \\
    P5-B   &  \textbf{2.1225} &  \textbf{8.4205} & \textbf{1.6676}  & \textbf{7.5476} \\
    ChatGPT  &0.3121 & 4.7177 & 0.6924& 4.2557\\
    ChatGLM w/o SFT & 0.2472  & 4.3544 & 0.6306 & 3.2348   \\
    LLaMA w/o SFT &0.4306&3.9954&0.8280&1.9456\\
    Alpaca w/o SFT &0.2020&2.9902&0.4285&2.2188\\
    \bottomrule
    \end{tabular}}
    \label{tab:summarization}
\end{table}

\subsection{Off-the-shelf Results}
\label{Sec: off-the-shelf result}
In this study, we conducted an evaluation of commonly used off-the-shelf models including ChatGPT, ChatGLM, LLaMA, and Alpaca as recommender systems in LLMRec for both accuracy-based and explainability-based tasks. In rating prediction, ChatGPT demonstrated RMSE and MAE scores of 1.4913 and 1.2836, respectively, as depicted in Table \ref{tab:rating}. In direct recommendation, ChatGPT remained substantially below trained models such as P5, as shown in Table \ref{tab:direct}. Furthermore, ChatGLM, LLaMA, and Alpaca were unable to produce standard results following the prompt requirement without fine-tuning, hence, were not directly applicable to these accuracy-based tasks.
\begin{table}
\centering
\footnotesize
\caption{Performance comparison on rating prediction.}
\resizebox{.3\textwidth}{!}{%
\begin{tabular}{ccc}
\toprule
Methods & RMSE   & MAE  \\
\cmidrule{1-3}
MF    &  \bf1.1973 & \bf 0.9461   \\
MLP    & 1.3078  & 0.9597  \\
ChatGPT   &   1.4913   &  1.2836   \\
ChatGLM w/o SFT  & N/A  & N/A    \\
LLaMA w/o SFT  & N/A  & N/A    \\
Alpaca w/o SFT  & N/A  & N/A    \\
\bottomrule
\end{tabular}
}
\label{tab:rating}
\end{table}

In the context of explainability-based tasks, it is evident that off-the-shelf models are progressively closing the gap with trained models and, in some cases, superseding baseline algorithms based on certain metrics, as clearly demonstrated in Table \ref{tab:explanation} and Table \ref{tab:summarization}. For instance, in review summarization task, ChatGPT achieves 4.7177, 0.6924, and 4.2557 on ROUGE1, ROUGE2, and ROUGEL metrics, respectively, exceeding the performance of T0 and GPT-2 trained on Beauty dataset. This suggests that it is possible to gain an insightful understanding of users' interest with current items, even without acquiring their historical interaction information. 
We also present samples generated by different models in supplementary materials, showing that ChatGPT's explanations are clearer and more rational, while P5 tends to produce generic-style explanations and performs the worst. The same conclusion can be drawn for review summarization task, where ChatGPT is able to comprehend the reviews and generate precise summaries, rather than simply extracting a few keywords.
\begin{table}[t]
    \centering
    \caption{Performance comparison on sequential recommendation.}
    \resizebox{\columnwidth}{!}{%
    \begin{tabular}{ccccc} 
    \toprule
    Methods
              & HR@5   & NDCG@5 & HR@10  & NDCG@10   \\ 
    \midrule
    Caser     & 0.0205 & 0.0131 & 0.0347 & 0.0176    \\
    HGN       & 0.0325 & 0.0206 & 0.0512 & 0.0266    \\
    GRU4Rec   & 0.0164 & 0.0099 & 0.0283 & 0.0137    \\
    BERT4Rec  & 0.0203 & 0.0124 & 0.0347 & 0.0170    \\
    FDSA      & 0.0267 & 0.0163 & 0.0407 & 0.0208    \\
    SASRec    & 0.0387 & 0.0249 & 0.0605 & 0.0318    \\
    S$^3$-Rec & 0.0387 & 0.0244 & 0.0647 & 0.0327    \\
    P5-B      & \textbf{0.0493} & \textbf{0.0367} & \textbf{0.0645} & \textbf{0.0416}    \\
    ChatGLM w/o SFT &  N/A  &  N/A &  N/A&  N/A    \\
    LLaMA w/o SFT&  N/A  &  N/A &  N/A&  N/A \\
    Alpaca w/o SFT&  N/A  &  N/A &  N/A&  N/A\\
    \bottomrule
    \end{tabular}
    }
    \label{tab:sequential}
    \end{table}
\begin{table}[t]
    \centering
    \caption{Performance comparison on direct recommendation.}
    \resizebox{\columnwidth}{!}{%
    \begin{tabular}{ccccc} 
    \toprule
    Methods & HR@5           & NDCG@5         & HR@10  & NDCG@10         \\ 
    \cmidrule{1-5}
    BPR-MF               & 0.1426         & 0.0857         & 0.2573 & 0.1224          \\
    BPR-MLP              & 0.1392         & 0.0848         & 0.2542 & 0.1215          \\
    SimpleX              & \textbf{0.2247} & \textbf{0.1441} & \textbf{0.3090} & \textbf{0.1711}  \\ 
    P5-B                 & 0.1564         & 0.1096         & 0.2300 & 0.1332          \\
    ChatGPT &    0.0217 & 0.0111 & 0.0652 & 0.0252    \\
    ChatGLM w/o SFT &  N/A  &  N/A &  N/A&  N/A    \\
    LLaMA w/o SFT&  N/A  &  N/A &  N/A&  N/A \\
    Alpaca w/o SFT&  N/A  &  N/A &  N/A&  N/A \\
    \bottomrule
    \end{tabular}
    }
    \label{tab:direct}
\end{table}

\begin{figure*}[t]
  \centering
  \includegraphics[width=0.9\textwidth]{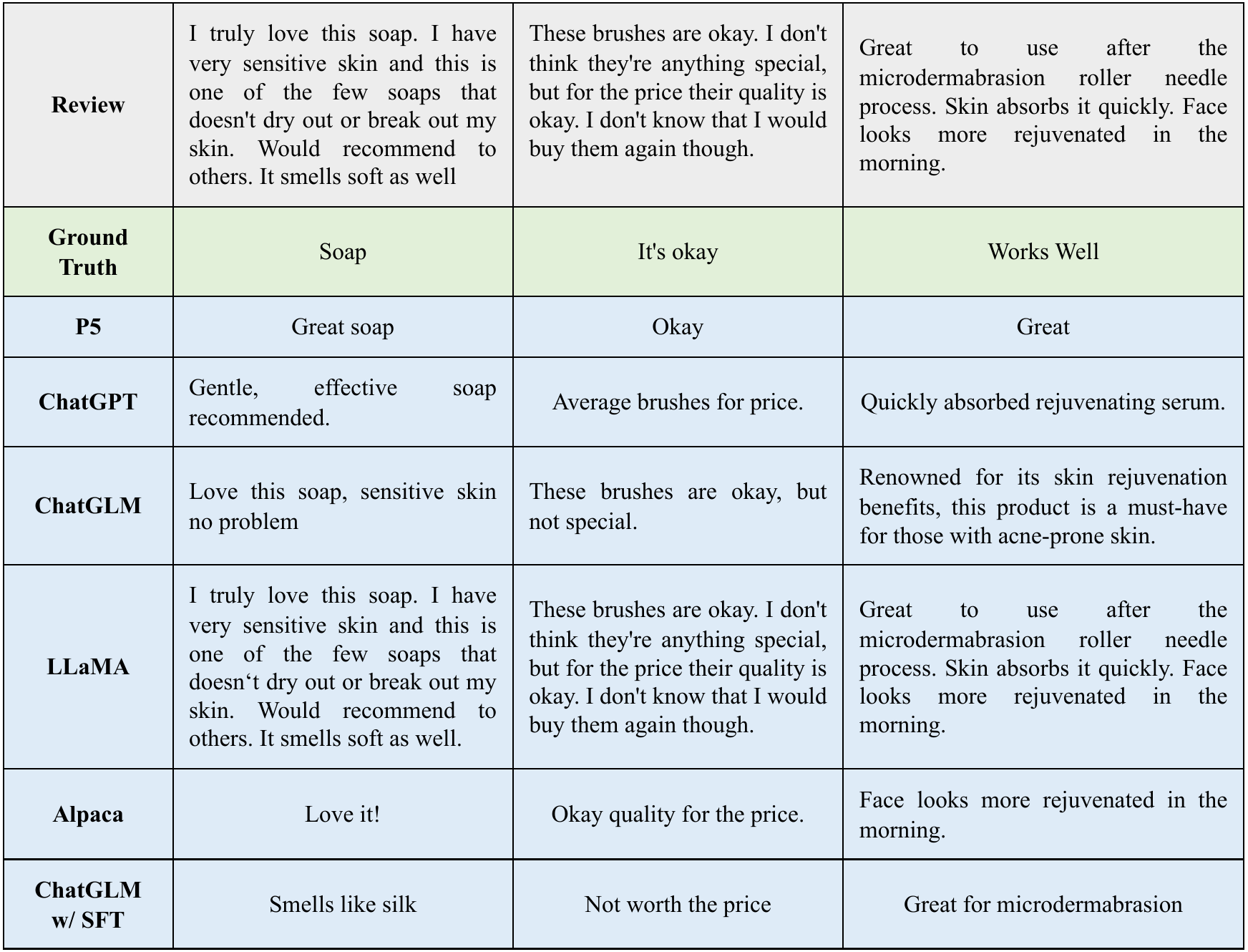}
  \caption{Example summarization results of different LLMs.}
  \label{fig:summarization_results}
\end{figure*}

\begin{table*}[t]
\centering
\caption{Fine-tuned LLMs' performance comparison on rating prediction, sequential recommendation and direct recommendation.}
\begin{adjustbox}{width=0.9\linewidth}
\begin{tabular}{ccc|cccc|ccccc}
\toprule
\multirow{2.5}{*}{Methods}  & \multicolumn{2}{c}{\textbf{Rating}} & \multicolumn{4}{c}{\textbf{Sequential}}  &\multicolumn{4}{c}{\textbf{Direct}}  \\
\cmidrule(lr){2-11}
 & RMSE   & MAE & HR@5   & NDCG@5& HR@10  & 
 NDCG@10  & HR@5  & NDCG@5  & HR@10  & 
 NDCG@10 \\
\cmidrule{1-11}
ChatGLM-6B w/ SFT & \bf 1.2912 & \bf 0.8945 & \bf 0.0403 & \bf 0.0349 & \bf 0.0455 & \bf 0.0366 & 0.0157 & 0.0105 & \bf 0.0205 & 0.0121 \\

\bottomrule
\end{tabular}
\end{adjustbox}
\vspace{-5pt}
\label{tab:sft-rating-sequential-direct}
\end{table*}

\begin{table*}[t]
\centering
\caption{Fine-tuned LLMs' performance comparison on explanation generation and review summarization (\%).}

\begin{adjustbox}{width=0.8\linewidth}
\begin{tabular}{ccccc|ccccc}
\toprule
\multirow{2.5}{*}{Methods}  & \multicolumn{4}{c}{\textbf{Explanation}}  &\multicolumn{4}{c}{\textbf{Review}}  \\
\cmidrule(lr){2-9}
 & BLUE4 & ROUGE1 & ROUGE2  & 
 ROUGEL & BLUE4  & ROUGE1 & ROUGE2 & ROUGEL \\
\cmidrule{1-9}
ChatGLM-6B w/ SFT & \bf 0.6766 & \bf 12.5927 & \bf 1.5564 & \bf 9.2806 & \bf 2.0254 & \bf 8.1003 & \bf 2.1351 & \bf 7.2126 \\
LLaMa-7B w/ SFT & 0.3006 & 11.3522 & 1.0774 & 5.1382 & 0.3464 & 5.7327 & 1.1382 & 2.7915\\
Alpaca w/ SFT & 0.3034 &11.7312 & 1.1018 & 5.3354 & 0.2752 & 5.1295 & 0.8269 & 2.4536   \\
\bottomrule
\end{tabular}
\end{adjustbox}
\label{tab:sft-explanation-review}
\end{table*}

\subsection{Fine-tuned Results}
From the section of off-the-shelf results, we could see that directly applying LLMs (\emph{e.g.}, ChatGPT) 
 in recommendation tasks results in inferior performance compared to baseline methods in certain recommendation tasks. In order to further explore the potential of utilizing LLMs in recommendation tasks, we conduct Supervised finetuning (SFT) on LLMs and test their performances on five recommendation tasks. Specifically, we construct prompts to test the original LLM's performance and use the training and testing datasets proposed by P5 \citep{geng2022recommendation} for SFT setting.

Table \ref{tab:sft-rating-sequential-direct} and Table \ref{tab:sft-explanation-review} show the fine-tuned results across the five recommendation tasks. From these results, we make the following observations:

\textbf{SFT helps LLMs align better with recommendation instructions.} For tasks that require formatted output (\emph{e.g.,} rating prediction), ChatGLM exhibits poor instruction compliance without fine-tuning, leading to unstructured outputs and, in some cases, an inability to complete the task. After supervised fine-tuning, its outputs can be restricted to the desired format. SFT also improves its automatic scores on natural-language tasks; for example, the ROUGEL score for explanation generation increases from 4.2388 to 9.2806.

% ChatGLM-6B, p-tuning v2, num_trainable params: 29360128
% P5-B: num_trainable params: 220000000
% fraction: 13%

\textbf{Comparision with P5} Although SFT makes LLMs such as ChatGLM capable to handle recommendation tasks like rating prediction or sequential recommendation, their performances are still far from stastified when compared with previous methods like P5 \cite{geng2022recommendation}. For example, P5-B get 0.0645 on HR@10 score but ChatGLM-6B only achieves 0.0455 after SFT. We attribute this results to the following reasons: First, the compared LLM methods have fewer amount of fine-tuning parameters. Take ChatGLM-6B as an example, we use P-tuning V2 \cite{liu2022ptuning} method to finetune the model, leading to 13\% of total trainable parameters compared to P5-B, which greatly limits the model's potential to bridge the gap between pretrained NLP tasks and finetuned recommendation tasks. Second, the amount of training data. Due to the limited calculating resouce, we could only finetune all compared LLM methods using one type of prompt for 10 Epochs for each kind of recommendation task, while P5 adopt a much larger training dataset that upsamples and merges all five task together using multiple kinds of prompt, resulting in about 50 times more training data. Third, the diversity in training data. As aforementioned, there are different types of task samples using multiple kinds of prompt in P5's training data, which greatly improves the diversity in training data and further strengthen the generization ability of the corresponding models. The multi-task training strategy in P5's setting could also improves the model's performance to some extent, which is also demonstrated in \cite{ruder2017overview}.

\textbf{Comparison with ChatGPT} 
The qualitative results demonstrate that ChatGPT performs well in explanation generation and review summarization. In the automatic evaluations, ChatGLM-6B with SFT outperforms ChatGPT in rating prediction, explanation generation, and review summarization, whereas ChatGPT performs better in direct recommendation. These findings suggest that SFT improves instruction compliance and the use of recommendation-domain information.
\subsection{Qualitative Evaluation}
In order to analyze the performance of LLMs in explainability-based tasks, we present the results of several models on review summarization tasks and explanation generation. Due to limited space, we only report the summarization results and readers can check explanation generation results in Appendix.3. As shown in Fig.\ref{fig:summarization_results}, while P5's summarization outcome has extracted some keywords, it has also disregarded crucial information that exists throughout the review. In contrast, large language models such as ChatGPT, ChatGLM, and Alpaca can generate more effective and meaningful summaries by comprehensively understanding and summarizing the reviews. However, LLaMA only duplicates the review and does not fulfill the task of summarization.
\section{Limitations}
\label{limiation}
We acknowledge several limitations in this study. Firstly, we have only evaluated specific Large Language Models (LLMs) such as ChatGPT, ChatGLM, LLaMA, and Alpaca. The findings may not be applicable to other models or future versions of these models. To address this, we plan to evaluate the recommendation performance of a broader range of LLMs in the future. 
Additionally, due to the expensive computational resources required for assessing LLM performance, we conducted experiments solely on the widely used Amazon Beauty dataset. However, due to different characteristics among datasets, we cannot guarantee that our findings will hold on other datasets. To mitigate this issue, we intend to conduct evaluations on more diverse datasets in the future. 
Furthermore, the study highlights the disagreement between objective and subjective evaluation results. In the future, we will explore objective metrics that align more closely with subjective evaluations.

\section{Conclusion}
 In this work, we present LLMRec, a general-purpose recommender system powered by LLMs. Using this system, we benchmark the performance of various LLMs in both off-the-shelf and supervised fine-tuning settings across five classical recommendation tasks.
 The results suggest that existing LLMs can perform competitively in rating prediction but remain limited in the reported accuracy-based evaluations. Notably, for explainable recommendation tasks, ChatGPT surpasses state-of-the-art methods in qualitative evaluations, indicating its potential in generating explanations and summaries for recommendation results. 
Our study contributes valuable insights into the strengths and limitations of LLMs in recommender systems. We hope our study can facilitate and encourage the development of more effective approaches for integrating LLMs into recommendation scenarios, bridging the semantic gap between language and user interests.

\appendix
\bibliography{aaai24}

@article{zhuo2023exploring,
  title={Exploring ai ethics of chatgpt: A diagnostic analysis},
  author={Zhuo, Terry Yue and Huang, Yujin and Chen, Chunyang and Xing, Zhenchang},
  journal={arXiv preprint arXiv:2301.12867},
  year={2023}
}

@article{liu2023evaluating,
  title={Evaluating the logical reasoning ability of chatgpt and gpt-4},
  author={Liu, Hanmeng and Ning, Ruoxi and Teng, Zhiyang and Liu, Jian and Zhou, Qiji and Zhang, Yue},
  journal={arXiv preprint arXiv:2304.03439},
  year={2023}
}

@article{li2023fairness,
  title={Fairness of ChatGPT},
  author={Li, Yunqi and Zhang, Yongfeng},
  journal={arXiv preprint arXiv:2305.18569},
  year={2023}
}

@article{radford2018improving,
  title={Improving language understanding by generative pre-training},
  author={Radford, Alec and Narasimhan, Karthik and Salimans, Tim and Sutskever, Ilya and others},
  year={2018},
  publisher={OpenAI}
}

@article{raffel2020exploring,
  title={Exploring the limits of transfer learning with a unified text-to-text transformer},
  author={Raffel, Colin and Shazeer, Noam and Roberts, Adam and Lee, Katherine and Narang, Sharan and Matena, Michael and Zhou, Yanqi and Li, Wei and Liu, Peter J},
  journal={The Journal of Machine Learning Research},
  volume={21},
  number={1},
  pages={5485--5551},
  year={2020},
  publisher={JMLRORG}
}

@inproceedings{du2022glm,
  title={GLM: General Language Model Pretraining with Autoregressive Blank Infilling},
  author={Du, Zhengxiao and Qian, Yujie and Liu, Xiao and Ding, Ming and Qiu, Jiezhong and Yang, Zhilin and Tang, Jie},
  booktitle={Proceedings of the 60th Annual Meeting of the Association for Computational Linguistics (Volume 1: Long Papers)},
  pages={320--335},
  year={2022}
}

@inproceedings{bert,
  author    = {Jacob Devlin and
               Ming{-}Wei Chang and
               Kenton Lee and
               Kristina Toutanova},
  title     = {{BERT:} Pre-training of Deep Bidirectional Transformers for Language
               Understanding},
  booktitle = {{NAACL-HLT} {(1)}},
  pages     = {4171--4186},
  publisher = {Association for Computational Linguistics},
  year      = {2019}
}

@article{2021GLM,
  title={GLM: General Language Model Pretraining with Autoregressive Blank Infilling},
  author={ Du, Z.  and  Qian, Y.  and  Liu, X.  and  Ding, M.  and  Qiu, J.  and  Yang, Z.  and  Tang, J. },
  year={2021},
}

@article{2017Get,
  title={Get To The Point: Summarization with Pointer-Generator Networks},
  author={ See, A.  and  Liu, P. J.  and  Manning, C. D. },
  year={2017},
}

@article{2019Fine,
  title={Fine-tune BERT for Extractive Summarization},
  author={ Liu, Y. },
  year={2019},
}

@article{2016A,
  title={A Persona-Based Neural Conversation Model},
  author={ Li, J.  and  Galley, M.  and  Brockett, C.  and  Spithourakis, G. P.  and  Gao, J.  and  Dolan, B. },
  journal={arXiv e-prints},
  year={2016},
}

@article{2016Towards,
  title={Towards End-to-End Reinforcement Learning of Dialogue Agents for Information Access},
  author={ Dhingra, B.  and  Li, L.  and  Li, X.  and  Gao, J.  and  Li, D. },
  year={2016},
}

@article{aharoni2019massively,
  title={Massively multilingual neural machine translation},
  author={Aharoni, Roee and Johnson, Melvin and Firat, Orhan},
  journal={arXiv preprint arXiv:1903.00089},
  year={2019}
}

@article{chen2018best,
  title={The best of both worlds: Combining recent advances in neural machine translation},
  author={Chen, Mia Xu and Firat, Orhan and Bapna, Ankur and Johnson, Melvin and Macherey, Wolfgang and Foster, George and Jones, Llion and Parmar, Niki and Schuster, Mike and Chen, Zhifeng and others},
  journal={arXiv preprint arXiv:1804.09849},
  year={2018}
}

@article{radford2019language,
  title={Language models are unsupervised multitask learners},
  author={Radford, Alec and Wu, Jeffrey and Child, Rewon and Luan, David and Amodei, Dario and Sutskever, Ilya and others},
  journal={OpenAI blog},
  volume={1},
  number={8},
  pages={9},
  year={2019}
}

@article{2003Journal,
  title={Journal of Machine Learning Research 3 (2003) 1137--1155 Submitted 4/02; Published 2/03 A Neural Probabilistic Language Model},
  author={ Bengio, Y.  and  Ducharme, R.  and  Vincent, P.  and  Jauvin, C.  and  Shawe-Taylor, J. },
  journal={JMLR.org},
  number={6},
  year={2003},
}

@article{2017Attention,
  title={Attention Is All You Need},
  author={ Vaswani, A.  and  Shazeer, N.  and  Parmar, N.  and  Uszkoreit, J.  and  Jones, L.  and  Gomez, A. N.  and  Kaiser, L.  and  Polosukhin, I. },
  journal={arXiv},
  year={2017},
}

@article{1997Long,
  title={Long Short-Term Memory},
  author={ Hochreiter, S.  and  Schmidhuber, J. },
  journal={Neural Computation},
  volume={9},
  number={8},
  pages={1735-1780},
  year={1997},
}

@article{gpt4,
  author    = {OpenAI},
  title     = {{GPT-4} Technical Report},
  journal   = {CoRR},
  volume    = {abs/2303.08774},
  year      = {2023}
}

@article{t5,
  author    = {Colin Raffel and
               Noam Shazeer and
               Adam Roberts and
               Katherine Lee and
               Sharan Narang and
               Michael Matena and
               Yanqi Zhou and
               Wei Li and
               Peter J. Liu},
  title     = {Exploring the Limits of Transfer Learning with a Unified Text-to-Text
               Transformer},
  journal   = {J. Mach. Learn. Res.},
  volume    = {21},
  pages     = {140:1--140:67},
  year      = {2020}
}

@article{m6,
  author    = {Zeyu Cui and
               Jianxin Ma and
               Chang Zhou and
               Jingren Zhou and
               Hongxia Yang},
  title     = {M6-Rec: Generative Pretrained Language Models are Open-Ended Recommender
               Systems},
  journal   = {CoRR},
  volume    = {abs/2205.08084},
  year      = {2022}
}

@inproceedings{
zhang2021language,
title={Language Models as Recommender Systems: Evaluations and Limitations},
author={Yuhui Zhang and HAO DING and Zeren Shui and Yifei Ma and James Zou and Anoop Deoras and Hao Wang},
booktitle={I (Still) Can't Believe It's Not Better! NeurIPS 2021 Workshop},
year={2021},
url={https://openreview.net/forum?id=hFx3fY7-m9b}
}

@article{chat-rec,
  title={Chat-REC: Towards Interactive and Explainable LLMs-Augmented Recommender System},
  author={Gao, Yunfan and Sheng, Tao and Xiang, Youlin and Xiong, Yun and Wang, Haofen and Zhang, Jiawei},
  journal={arXiv preprint arXiv:2303.14524},
  year={2023}
}

@inproceedings{zhou2020s3,
  title={S3-rec: Self-supervised learning for sequential recommendation with mutual information maximization},
  author={Zhou, Kun and Wang, Hui and Zhao, Wayne Xin and Zhu, Yutao and Wang, Sirui and Zhang, Fuzheng and Wang, Zhongyuan and Wen, Ji-Rong},
  booktitle={Proceedings of the 29th ACM international conference on information \& knowledge management},
  pages={1893--1902},
  year={2020}
}

@inproceedings{geng2022recommendation,
  title={Recommendation as language processing (rlp): A unified pretrain, personalized prompt \& predict paradigm (p5)},
  author={Geng, Shijie and Liu, Shuchang and Fu, Zuohui and Ge, Yingqiang and Zhang, Yongfeng},
  booktitle={Proceedings of the 16th ACM Conference on Recommender Systems},
  pages={299--315},
  year={2022}
}

@article{koren2009matrix,
  title={Matrix factorization techniques for recommender systems},
  author={Koren, Yehuda and Bell, Robert and Volinsky, Chris},
  journal={Computer},
  volume={42},
  number={8},
  pages={30--37},
  year={2009},
  publisher={IEEE}
}

@inproceedings{cheng2016wide,
  title={Wide \& deep learning for recommender systems},
  author={Cheng, Heng-Tze and Koc, Levent and Harmsen, Jeremiah and Shaked, Tal and Chandra, Tushar and Aradhye, Hrishi and Anderson, Glen and Corrado, Greg and Chai, Wei and Ispir, Mustafa and others},
  booktitle={Proceedings of the 1st workshop on deep learning for recommender systems},
  pages={7--10},
  year={2016}
}

@article{rendle2012bpr,
  title={BPR: Bayesian personalized ranking from implicit feedback},
  author={Rendle, Steffen and Freudenthaler, Christoph and Gantner, Zeno and Schmidt-Thieme, Lars},
  journal={arXiv preprint arXiv:1205.2618},
  year={2012}
}

@inproceedings{mao2021simplex,
  title={SimpleX: A simple and strong baseline for collaborative filtering},
  author={Mao, Kelong and Zhu, Jieming and Wang, Jinpeng and Dai, Quanyu and Dong, Zhenhua and Xiao, Xi and He, Xiuqiang},
  booktitle={Proceedings of the 30th ACM International Conference on Information \& Knowledge Management},
  pages={1243--1252},
  year={2021}
}

@inproceedings{tang2018personalized,
  title={Personalized top-n sequential recommendation via convolutional sequence embedding},
  author={Tang, Jiaxi and Wang, Ke},
  booktitle={Proceedings of the eleventh ACM international conference on web search and data mining},
  pages={565--573},
  year={2018}
}

@inproceedings{ma2019hierarchical,
  title={Hierarchical gating networks for sequential recommendation},
  author={Ma, Chen and Kang, Peng and Liu, Xue},
  booktitle={Proceedings of the 25th ACM SIGKDD international conference on knowledge discovery \& data mining},
  pages={825--833},
  year={2019}
}

@article{hidasi2015session,
  title={Session-based recommendations with recurrent neural networks},
  author={Hidasi, Bal{\'a}zs and Karatzoglou, Alexandros and Baltrunas, Linas and Tikk, Domonkos},
  journal={arXiv preprint arXiv:1511.06939},
  year={2015}
}

@inproceedings{sun2019bert4rec,
  title={BERT4Rec: Sequential recommendation with bidirectional encoder representations from transformer},
  author={Sun, Fei and Liu, Jun and Wu, Jian and Pei, Changhua and Lin, Xiao and Ou, Wenwu and Jiang, Peng},
  booktitle={Proceedings of the 28th ACM international conference on information and knowledge management},
  pages={1441--1450},
  year={2019}
}

@inproceedings{zhang2019feature,
  title={Feature-level Deeper Self-Attention Network for Sequential Recommendation.},
  author={Zhang, Tingting and Zhao, Pengpeng and Liu, Yanchi and Sheng, Victor S and Xu, Jiajie and Wang, Deqing and Liu, Guanfeng and Zhou, Xiaofang and others},
  booktitle={IJCAI},
  pages={4320--4326},
  year={2019}
}

@inproceedings{kang2018self,
  title={Self-attentive sequential recommendation},
  author={Kang, Wang-Cheng and McAuley, Julian},
  booktitle={2018 IEEE international conference on data mining (ICDM)},
  pages={197--206},
  year={2018},
  organization={IEEE}
}

@inproceedings{dong2017learning,
  title={Learning to generate product reviews from attributes},
  author={Dong, Li and Huang, Shaohan and Wei, Furu and Lapata, Mirella and Zhou, Ming and Xu, Ke},
  booktitle={Proceedings of the 15th Conference of the European Chapter of the Association for Computational Linguistics: Volume 1, Long Papers},
  pages={623--632},
  year={2017}
}

@inproceedings{li2017neural,
  title={Neural rating regression with abstractive tips generation for recommendation},
  author={Li, Piji and Wang, Zihao and Ren, Zhaochun and Bing, Lidong and Lam, Wai},
  booktitle={Proceedings of the 40th International ACM SIGIR conference on Research and Development in Information Retrieval},
  pages={345--354},
  year={2017}
}

@article{li2021personalized,
  title={Personalized transformer for explainable recommendation},
  author={Li, Lei and Zhang, Yongfeng and Chen, Li},
  journal={arXiv preprint arXiv:2105.11601},
  year={2021}
}

@article{sanh2021multitask,
  title={Multitask prompted training enables zero-shot task generalization},
  author={Sanh, Victor and Webson, Albert and Raffel, Colin and Bach, Stephen H and Sutawika, Lintang and Alyafeai, Zaid and Chaffin, Antoine and Stiegler, Arnaud and Scao, Teven Le and Raja, Arun and others},
  journal={arXiv preprint arXiv:2110.08207},
  year={2021}
}

@inproceedings{sun2022revisiting,
  title={Revisiting Bundle Recommendation: Datasets, Tasks, Challenges and Opportunities for Intent-aware Product Bundling},
  author={Sun, Zhu and Yang, Jie and Feng, Kaidong and Fang, Hui and Qu, Xinghua and Ong, Yew Soon},
  booktitle={Proceedings of the 45th International ACM SIGIR Conference on Research and Development in Information Retrieval},
  pages={2900--2911},
  year={2022}
}

@article{liu2022ecommerce,
  title={An ecommerce recommendation algorithm based on link prediction},
  author={Liu, Guoguang},
  journal={Alexandria Engineering Journal},
  volume={61},
  number={1},
  pages={905--910},
  year={2022},
  publisher={Elsevier}
}

@inproceedings{tsagkias2021challenges,
  title={Challenges and research opportunities in ecommerce search and recommendations},
  author={Tsagkias, Manos and King, Tracy Holloway and Kallumadi, Surya and Murdock, Vanessa and de Rijke, Maarten},
  booktitle={ACM Sigir Forum},
  volume={54},
  number={1},
  pages={1--23},
  year={2021},
  organization={ACM New York, NY, USA}
}

@inproceedings{wei2019mmgcn,
  title={MMGCN: Multi-modal graph convolution network for personalized recommendation of micro-video},
  author={Wei, Yinwei and Wang, Xiang and Nie, Liqiang and He, Xiangnan and Hong, Richang and Chua, Tat-Seng},
  booktitle={Proceedings of the 27th ACM international conference on multimedia},
  pages={1437--1445},
  year={2019}
}

@inproceedings{zhao2019recommending,
  title={Recommending what video to watch next: a multitask ranking system},
  author={Zhao, Zhe and Hong, Lichan and Wei, Li and Chen, Jilin and Nath, Aniruddh and Andrews, Shawn and Kumthekar, Aditee and Sathiamoorthy, Maheswaran and Yi, Xinyang and Chi, Ed},
  booktitle={Proceedings of the 13th ACM Conference on Recommender Systems},
  pages={43--51},
  year={2019}
}

@inproceedings{papadamou2022just,
  title={“It is just a flu”: Assessing the Effect of Watch History on YouTube’s Pseudoscientific Video Recommendations},
  author={Papadamou, Kostantinos and Zannettou, Savvas and Blackburn, Jeremy and De Cristofaro, Emiliano and Stringhini, Gianluca and Sirivianos, Michael},
  booktitle={Proceedings of the international AAAI conference on web and social media},
  volume={16},
  pages={723--734},
  year={2022}
}

@inproceedings{wu2022feedrec,
  title={Feedrec: News feed recommendation with various user feedbacks},
  author={Wu, Chuhan and Wu, Fangzhao and Qi, Tao and Liu, Qi and Tian, Xuan and Li, Jie and He, Wei and Huang, Yongfeng and Xie, Xing},
  booktitle={Proceedings of the ACM Web Conference 2022},
  pages={2088--2097},
  year={2022}
}

@inproceedings{wu2020mind,
  title={Mind: A large-scale dataset for news recommendation},
  author={Wu, Fangzhao and Qiao, Ying and Chen, Jiun-Hung and Wu, Chuhan and Qi, Tao and Lian, Jianxun and Liu, Danyang and Xie, Xing and Gao, Jianfeng and Wu, Winnie and others},
  booktitle={Proceedings of the 58th Annual Meeting of the Association for Computational Linguistics},
  pages={3597--3606},
  year={2020}
}

@inproceedings{wu2019npa,
  title={NPA: neural news recommendation with personalized attention},
  author={Wu, Chuhan and Wu, Fangzhao and An, Mingxiao and Huang, Jianqiang and Huang, Yongfeng and Xie, Xing},
  booktitle={Proceedings of the 25th ACM SIGKDD international conference on knowledge discovery \& data mining},
  pages={2576--2584},
  year={2019}
}

@inproceedings{kowald2020unfairness,
  title={The unfairness of popularity bias in music recommendation: A reproducibility study},
  author={Kowald, Dominik and Schedl, Markus and Lex, Elisabeth},
  booktitle={Advances in Information Retrieval: 42nd European Conference on IR Research, ECIR 2020, Lisbon, Portugal, April 14--17, 2020, Proceedings, Part II 42},
  pages={35--42},
  year={2020},
  organization={Springer}
}

@inproceedings{singh2022novel,
  title={A Novel Deep Neural-based Music Recommendation Method considering User and Song Data},
  author={Singh, Jagendra and Sajid, Mohammad and Yadav, Chandra Shekhar and Singh, Shashank Sheshar and Saini, Manthan},
  booktitle={2022 6th International Conference on Trends in Electronics and Informatics (ICOEI)},
  pages={1--7},
  year={2022},
  organization={IEEE}
}

@article{dai2023chataug,
  title={Chataug: Leveraging chatgpt for text data augmentation},
  author={Dai, Haixing and Liu, Zhengliang and Liao, Wenxiong and Huang, Xiaoke and Wu, Zihao and Zhao, Lin and Liu, Wei and Liu, Ninghao and Li, Sheng and Zhu, Dajiang and others},
  journal={arXiv preprint arXiv:2302.13007},
  year={2023}
}

@article{jiao2023chatgpt,
  title={Is ChatGPT a good translator? A preliminary study},
  author={Jiao, Wenxiang and Wang, Wenxuan and Huang, Jen-tse and Wang, Xing and Tu, Zhaopeng},
  journal={arXiv preprint arXiv:2301.08745},
  year={2023}
}

@article{bang2023multitask,
  title={A multitask, multilingual, multimodal evaluation of chatgpt on reasoning, hallucination, and interactivity},
  author={Bang, Yejin and Cahyawijaya, Samuel and Lee, Nayeon and Dai, Wenliang and Su, Dan and Wilie, Bryan and Lovenia, Holy and Ji, Ziwei and Yu, Tiezheng and Chung, Willy and others},
  journal={arXiv preprint arXiv:2302.04023},
  year={2023}
}

@inproceedings{xie2022decoupled,
  title={Decoupled side information fusion for sequential recommendation},
  author={Xie, Yueqi and Zhou, Peilin and Kim, Sunghun},
  booktitle={Proceedings of the 45th International ACM SIGIR Conference on Research and Development in Information Retrieval},
  pages={1611--1621},
  year={2022}
}

@article{zeng2022greenplm,
  title={GreenPLM: Cross-lingual pre-trained language models conversion with (almost) no cost},
  author={Zeng, Qingcheng and Garay, Lucas and Zhou, Peilin and Chong, Dading and Hua, Yining and Wu, Jiageng and Pan, Yikang and Zhou, Han and Yang, Jie},
  journal={arXiv preprint arXiv:2211.06993},
  year={2022}
}

@article{zhang2023llama,
  title={Llama-adapter: Efficient fine-tuning of language models with zero-init attention},
  author={Zhang, Renrui and Han, Jiaming and Zhou, Aojun and Hu, Xiangfei and Yan, Shilin and Lu, Pan and Li, Hongsheng and Gao, Peng and Qiao, Yu},
  journal={arXiv preprint arXiv:2303.16199},
  year={2023}
}

@article{hu2021lora,
  title={Lora: Low-rank adaptation of large language models},
  author={Hu, Edward J and Shen, Yelong and Wallis, Phillip and Allen-Zhu, Zeyuan and Li, Yuanzhi and Wang, Shean and Wang, Lu and Chen, Weizhu},
  journal={arXiv preprint arXiv:2106.09685},
  year={2021}
}

@misc{liu2022ptuning,
      title={P-Tuning v2: Prompt Tuning Can Be Comparable to Fine-tuning Universally Across Scales and Tasks}, 
      author={Xiao Liu and Kaixuan Ji and Yicheng Fu and Weng Lam Tam and Zhengxiao Du and Zhilin Yang and Jie Tang},
      year={2022},
      eprint={2110.07602},
      archivePrefix={arXiv},
      primaryClass={cs.CL}
}

@article{ruder2017overview,
  title={An overview of multi-task learning in deep neural networks},
  author={Ruder, Sebastian},
  journal={arXiv preprint arXiv:1706.05098},
  year={2017}
}

@misc{alpaca,
  author = {Rohan Taori and Ishaan Gulrajani and Tianyi Zhang and Yann Dubois and Xuechen Li and Carlos Guestrin and Percy Liang and Tatsunori B. Hashimoto },
  title = {Stanford Alpaca: An Instruction-following LLaMA model},
  year = {2023},
  publisher = {GitHub},
  journal = {GitHub repository},
  howpublished = {\url{https://github.com/tatsu-lab/stanford_alpaca}},
}

@misc{liu2023chatgpt,
      title={Is ChatGPT a Good Recommender? A Preliminary Study}, 
      author={Junling Liu and Chao Liu and Peilin Zhou and Renjie Lv and Kang Zhou and Yan Zhang},
      year={2023},
      eprint={2304.10149},
      archivePrefix={arXiv},
      primaryClass={cs.IR}
}

\clearpage

\section{Appendix.1 Dataset Statistics and Splits}
\textit{Basic Statistics.} We conducted numerical and human evaluations on the real-world Amazon recommendation dataset, specifically focusing on the Beauty category. This dataset exhibits the following basic statistics: there are 22,363 users, 12,101 items, 198,502 reviews, and 198,502 user-item actions. On average, each user performs 8.9 actions, while each item receives an average of 16.4 actions. The sparsity of this dataset is 99.93\%.

\textit{Dataset Splits.} Following P5 \citep{geng2022recommendation}, we adopt different data partitioning strategies for different tasks. Specifically, for rating, explanation, and review tasks, we divide the Amazon Beauty dataset into training, validation, and testing sets using a ratio of 8:1:1. We ensure that each user and item has at least one instance included in the training set. To obtain ground-truth explanations, Sentires toolkit~\footnote{https://github.com/evison/Sentires} is first utilized to extract item feature words from the reviews and then sentences that comment on one or multiple item features are regarded as explanations of user preferences. For the sequential recommendation task, we employ a leave-one-out strategy to split the dataset: for each interaction sequence, the last item is treated as the test data, the item before the last one as the validation data, and the remaining data for training. In the direct recommendation task, the training set is consistent with the training split of the sequential recommendation task to avoid data leakage issues during pretraining.  

\section{Appendix.2 Implementation Details}
The configuration and hyper-parameters of the LLMs used in our work have been presented in the Sec 4.2 of our manuscript.
To ensure a fair comparison with P5, we maintain consistency with their dataset splits for each task. Consequently, in the main table of the manuscript, we directly inherit the performance results of the baseline and P5 reported in \citep{geng2022recommendation}.

For reproducibility, we provide the following information on the hardware used for training and testing the LLMs:
\begin{itemize}
  \item CPU: AMD EPYC 7713
  \item GPU: NVIDIA A100 SXM4 80GB
  \item RAM: 1024GB DDR4 ECC
\end{itemize}
All the experiments were conducted in the following environment:
\begin{itemize}
  \item Operating System: Ubuntu 20.04
  \item Programming Language: Python 3.10
  \item Libraries and Frameworks: Transformers 4.30.0, PEFT 0.4.0.dev0, PyTorch 2.0.1
\end{itemize}

Please ensure that the exact versions of these libraries and frameworks are installed to guarantee reproducibility. The full list of dependencies can be found in the requirements.txt file located in our code appendix.

\section{Appendix.3 In-depth Analysis}
\subsubsection{Case study of N/A results}

In Table ~3-5 of the manuscript, we denote the performance of off-the-shelf LLM models including ChatGLM, LLAMA, and Alpaca, as "N/A" in rating prediction, sequential recommendation, and direct recommendation tasks. The main reason for this is that, as shown in Supplement Fig~\ref{fig:na}, the output of LLAMA and Alpaca cannot be parsed into scores or item title using pre-defined rules. While ChatGLM is capable of generating some seemingly reasonable results in sequential recommendation tasks, the generated titles are mostly either present in the prompts or fake ones that do not exist in the dataset, leading to very low accuracy metrics. Therefore, we also mark it as N/A. To summarize, these "N/A" results indicate that LLM models with smaller parameter sizes are unable to complete accuracy-driven recommendation tasks without fine-tuning on recommendation datasets.

\subsubsection{Case study of explanation and summarization results}
In order to analyze the performance of LLMs in explainability-based tasks, we present the results of several models on explanation generation and review summarization tasks. P5 and general LLMs have distinct design goals and diverse applications as language models. P5 aims to generate explanatory language that resembles known texts and therefore places emphasis on learning text structure and grammar rules during training, resulting in more standardized generated output, as illustrated in Supplement Fig.\ref{fig:explanation_results}. In contrast, ChatGPT prioritizes language interaction and diversity. Its principal application is to simulate human conversation, and thus it considers multiple factors, such as context, emotion, and logic, to better express human thinking and language habits. This design approach leads to more diverse and creative text generated by ChatGPT. ChatGLM is also capable of achieving commendable results. However, LLaMA is prone to more noticeable issues with generating duplicates, while some explanations generated by Alpaca may be incomplete. With the application of Parameter Efficient Fine Tuning, it can be observed that ChatGLM is capable of achieving outcomes that are more closely aligned with the ground truth.

Similar to the explanation generation task, while P5's summarization outcome has extracted some keywords, it has also disregarded crucial information that exists throughout the review. In contrast, language models such as ChatGPT, ChatGLM, and Alpaca can generate more effective and meaningful summaries by comprehensively understanding and summarizing the reviews, as illustrated in Manuscript Fig.3. However, LLaMA only duplicates the review and does not fulfill the task of summarization.

\begin{figure*}[t]
  \centering
  \includegraphics[width=1.0\textwidth]{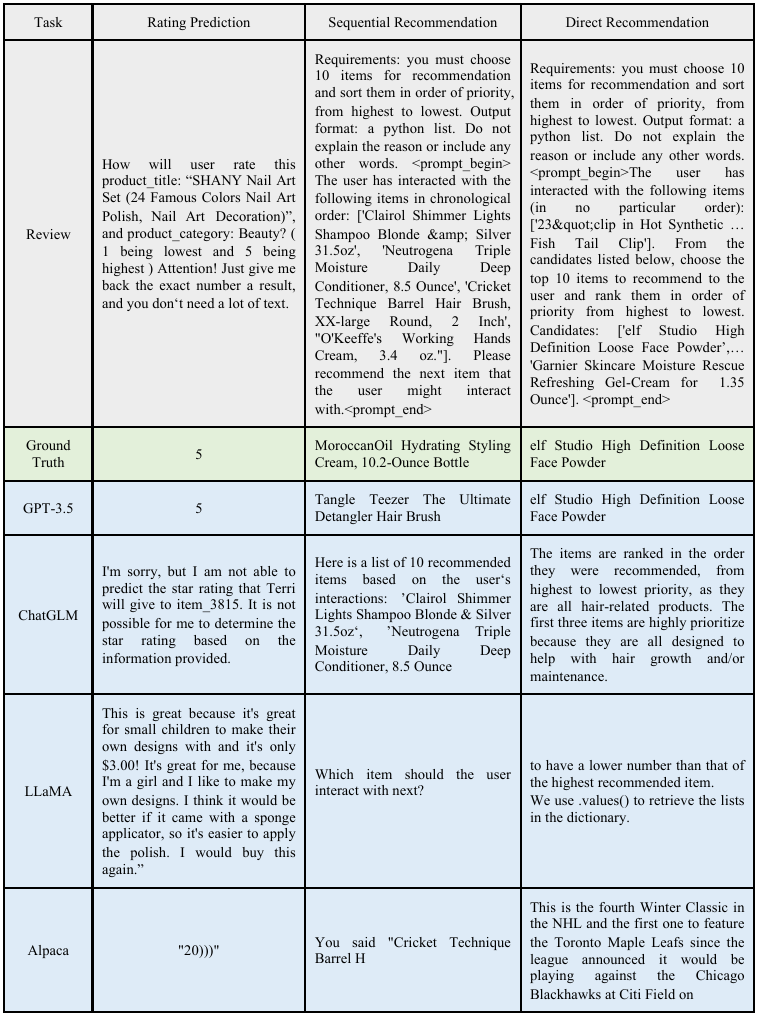}
  \caption{Case study of N/A results.}
  \label{fig:na}
\end{figure*}

\begin{figure*}[t]
  \centering
  \includegraphics[width=1\textwidth]{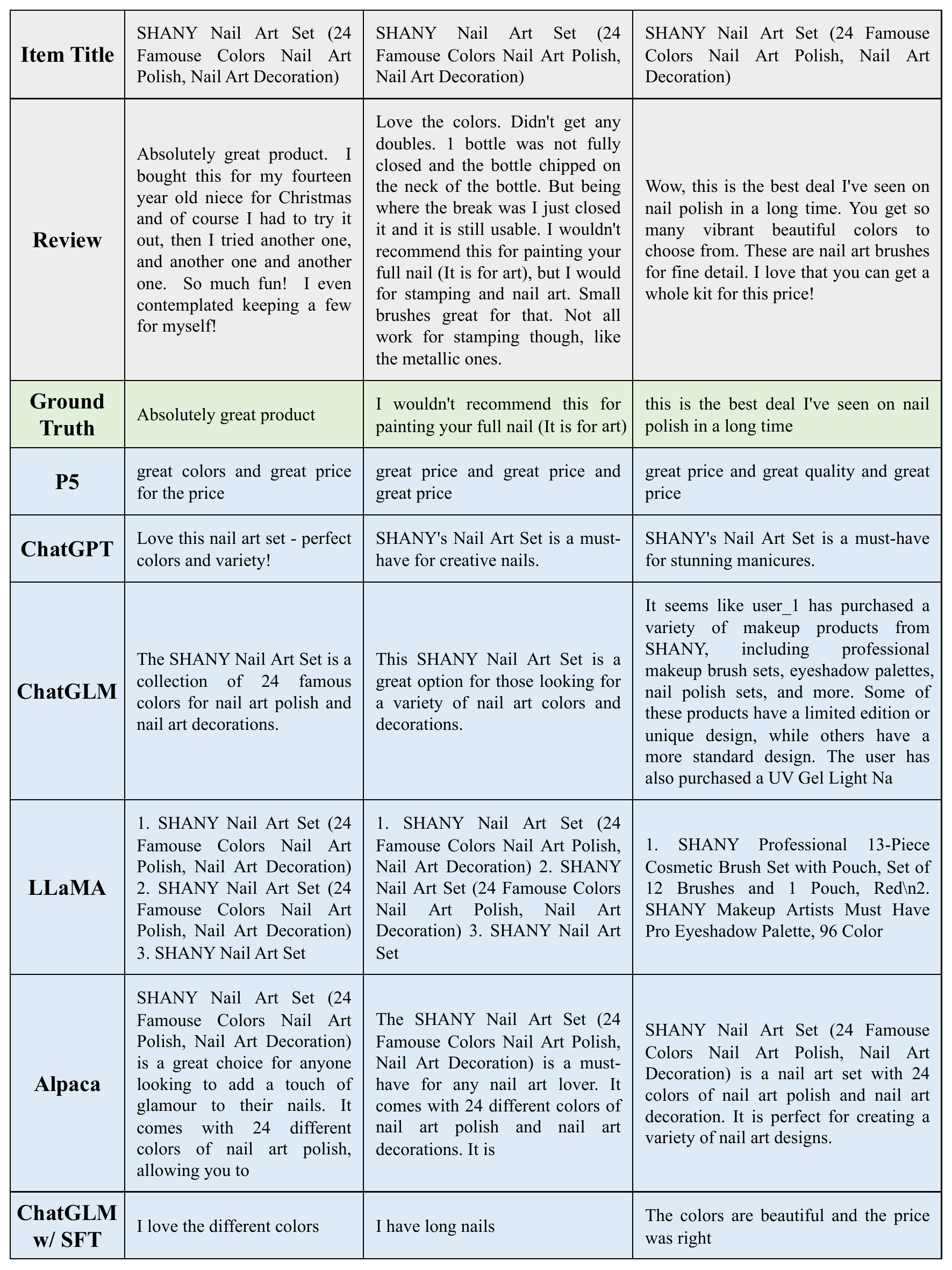}
  \caption{Example explanation results of different LLMs.}
  \label{fig:explanation_results}
\end{figure*}

\clearpage

\end{document}